\documentclass[twocolumn,showpacs,aps,apl]{revtex4}
\usepackage{graphicx}
\usepackage{dcolumn}
\usepackage{bm}

\begin{document}

\title{Effective tuning of exciton polarization splitting in coupled quantum dots}

\author{Jia-Lin Zhu}
\email[Author to whom correspondence should be addressed; electronic address:
]{zjl-dmp@tsinghua.edu.cn}
\affiliation{%
Department of Physics, Tsinghua University, Beijing 100084, People's Republic
of China}
\author{Dong Xu}
\affiliation{%
Department of Physics, Tsinghua University, Beijing 100084, People's Republic
of China}
\date{\today}

\begin{abstract}
The polarization splitting of the exciton ground state in two laterally coupled
quantum dots under an in-plane electric field is investigated and its effective
tuning is designed. It is found that there are significant Stark effect and
anticrossing in energy levels. Due to coupling between inter- and intra-dot
states, the absolute value of polarization splitting is significantly reduced,
and it could be tuned to zero by the electric field for proper inter-dot
separations. Our scheme is interesting for the research on the quantum
dots-based entangled-photon source.
\end{abstract}
\pacs{71.35.-y, 78.67.-n, 71.70.Gm, 78.55.Cr}
\maketitle

Semiconductor quantum dots (QDs) have been demonstrated as one of candidates
for both the single-photon and entangled two-photon sources, which make them
very attractive for applications in the fields of quantum cryptography, quantum
teleportation and quantum
computation.~\cite{michler1,Akopian,Santori1,Santori2,Stevenson} In the first
proposal for a QD-based source of polarization entangled photon pairs, a
necessary condition is that the intermediate exciton states for the biexciton
radiative decay are energetically degenerate.~\cite{Benson} However, the
monoexciton ground states in III-V semiconductor QDs are split by the
anisotropic electron-hole exchange interaction since QDs tend to be elongated
along the $[\bar{1}10]$ crystal
axis.~\cite{Gammon,Bayer,Ivchenko,Takagahara,Bester1,Seguin} Recently, it has
been demonstrated that a magnetic field in a Voigt-configuration could be used
to obtain degenerate exciton ground states.~\cite{Stevenson,Stevenson2}
Nevertheless, it is difficult to obtain degenerate exciton states in a single
QD by an electric field.~\cite{Kowalik} In this letter, we study the
polarization splitting of exciton ground states in two laterally coupled
quantum dots (CQDs) and find that it could be tuned to zero by an in-plane
electric field for proper inter-dot couplings.

Recently, with the development of high-quality InGaAs QD structures, it is
possible to fabricate laterally CQDs.~\cite{Beirne} Significant Stark effects
induced by an in-plane electric field have been observed in the experiment. The
exciton ground states might exhibit fine structures induced by the coherent
tunnel-coupling between the two dots. In this letter, we calculate exciton
polarization splitting in laterally CQDs under an in-plane electric field.
According to a recent empirical tight-binding calculation,~\cite{Sheng} the
heavy-hole component is dominant in the hole ground state of flat InGaAs QDs.
Therefore, it is reasonable that the light-hole and spin-orbit-split $J=1/2$
valence bands could be neglected in our calculations since we mainly focus on
the polarization splitting of the exciton ground states. Thus the exciton state
is composted of 4 combinations of the valence band and the conduction band,
i.e., $|X\rangle=\sum_{m,s}\sum_{r_e,r_h}
\psi_{ms}(r_e,r_h)a^{\dag}_{c_sr_e}a_{v_mr_h}|0\rangle$ where the Wannier
function representation of the creation and annihilation operators is used, $m$
and $s$ are the $z$-component of the angular momentum of the heavy-hole valence
band and the conduction band, respectively, and $\psi_{ms}(r_e,r_h)$ is the
exciton envelope function. The eigenvalue equation for $\psi_{ms}$ is given as
{\setlength\arraycolsep{2pt}
\begin{eqnarray}\label{eigenvalue equation}
\sum_{m's'r'_er'_h}[H_1&+&V_{\mathrm{ex}}(c_sr_e,v_{m'}r'_h;c_{s'}r'_e,v_mr_h)]\psi_{m's'}(r'_e,r'_h)
\nonumber\\&&=E\psi_{ms}(r_e,r_h),
\end{eqnarray}}
with
\begin{eqnarray}\label{spin-independent Hamiltonian}
H_1=\delta_{r_er'_e}\delta_{r_hr'_h}\delta_{s's}\delta_{m'm}[\frac{p^2_e}{2m_e}+U_e(r_e)
\nonumber\\+\frac{p^2_h}{2m_h}+U_h(r_h)+e\mathrm{F}\cdot(\mathrm{r}_h-\mathrm{r}_e)-\frac{e^2}{\epsilon|\mathrm{r}_e-\mathrm{r}_h|}],
\end{eqnarray}
where $U_h$ ($U_e$) is the confinement potential for the valence (conduction)
band electron and $\mathrm{F}$ is an in-plane electric field.
$V_{\mathrm{ex}}$, the electron-hole exchange interaction, splits the exciton
states into dark and bright doublets.~\cite{Takagahara} The bright doublets
typically consist of two linearly polarized states, the splitting of which
(polarization splitting) is about tens of $\mu$eV and mainly contributed from
long-range exchange interaction.~\cite{Bayer,Takagahara} Similar to the
assumption in Ref.~\onlinecite{Ivchenko}, an in-plane anisotropic potential is
used to model a single QD, and two laterally coupled dots are aligned along the
$x$-axis
\begin{eqnarray}\label{potential}
&&U_{e(h)}=\nu_{e(h)}[\theta(\frac{b_0}{2}-|y_{e(h)}|)\theta(\frac{a_0}{2}-|x_{e(h)}+\frac{d+a_0}{2}|)
\nonumber\\&&{}+\theta(\frac{b_1}{2}-|y_{e(h)}|)\theta(\frac{a_1}{2}-|x_{e(h)}-\frac{d+a_1}{2}|)]
\end{eqnarray}
where $a_i$ and $b_i$ are the lateral sizes of the $i$th dot, $d$ is the
inter-dot separation, and $\nu_{e}$ ($\nu_{h}$) is the conduction (heavy-hole
valence) band offset. Whether the geometric shape of single QDs is rectangular
or elliptic will not change the qualitative results of this letter. The
computational procedure is that eigenfunctions of spin-independent $H_1$ is
firstly calculated using diagonalization method, and then the polarization
splitting is obtained by calculating the matrix elements of $V_{\mathrm{ex}}$
in the basis of eigenfunctions of $H_1$.

Considering the size fluctuation in real QDs samples, we investigate two
coupled nonidentical QDs with sizes of $a_0=a_1=16$ nm, $b_0=19$ nm, $b_1=20$
nm. The inter-dot separation $d$ could be adjusted in the range of several
nanometers. In coupled nonidentical QDs, the energy difference between exciton
ground-state energies in two isolated nonidentical QDs are several meV, which
is much larger than the typical exchange-interaction energy. However, the
energy difference is comparable with the electron and hole tunnelling energies
in the strong coupling region of CQDs. Thus, the coupling between two QDs is
predominantly determined by the electron and hole tunnelling as well as the
electron-hole Coulomb correlation.

\begin{figure}
\begin{center}
\includegraphics*[angle=0,width=0.43\textwidth]{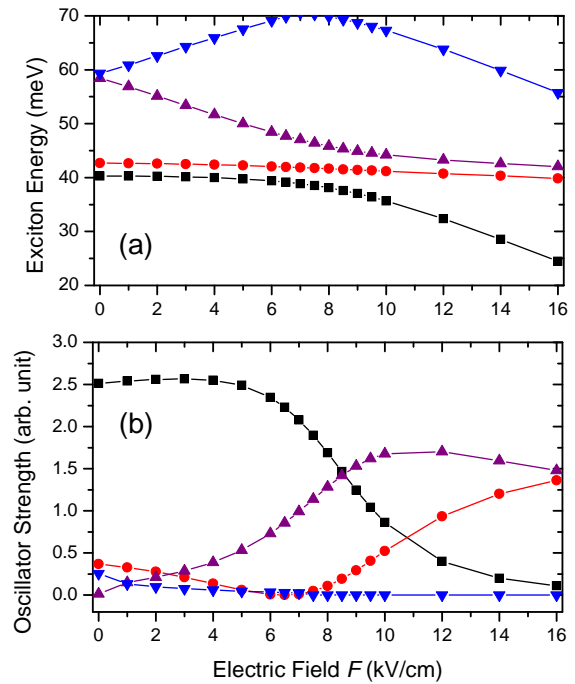}
\caption{(Color online) (a) The four-lowest energy levels of exciton and (b)
corresponding oscillator strength in the CQDs with $d=4.0$ nm as functions of
the in-plane electric field.}\label{FIG:fs cqds}
\end{center}
\end{figure}

Before giving the dependence of the polarization splitting on the in-plane
electric field, we investigate the exciton states without the exchange
interaction in CQDs. Fig.~\ref{FIG:fs cqds}(a) shows the numerically-calculated
lowest four energy levels of exciton in CQDs with $d=4.0$ nm, as functions of
the electric field along the $x$-axis. It can be clearly seen that there are
significant Stark effect and anticrossing in energy levels. Fig.~\ref{FIG:fs
cqds}(b) shows the corresponding oscillator strength of the lowest four levels
as functions of $\mathrm{F}$. At zero electric field, the first and second
energy levels are optically active while the third energy level is dark state.
However, as the electric field is applied, the third levels becomes optically
active. Oscillator strength of the second level initially decreases and then
increases significantly with increasing electric field. Moreover, the ground
state becomes optically inactive at larger electric field.

\begin{figure}
\begin{center}
\includegraphics*[angle=0,width=0.4\textwidth]{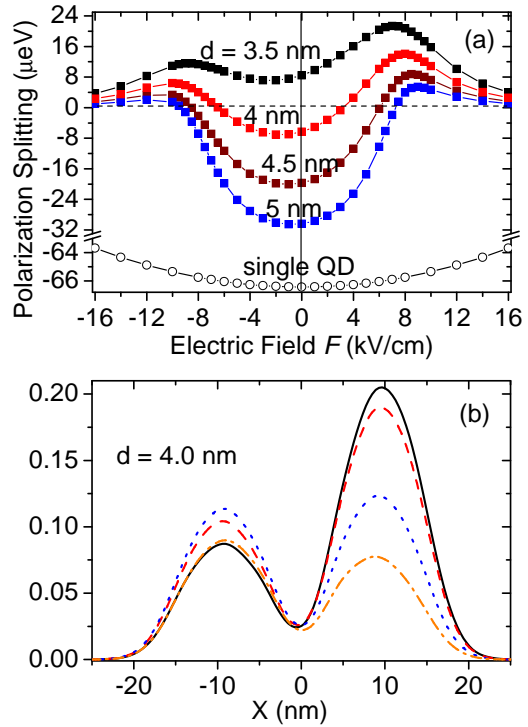}
\caption{(Color online) (a) Polarization splitting of the exciton ground state
as a function of the in-plane electric field in the CQDs with $d=3.5$, 4.0,
4.5, and 5.0 nm, and in a single QD, respectively. (b) $\psi(r_e,r_h=r_e)$ of
the exciton ground state alone $y=0$ axis in the CQDs with $d=4.0$ nm for $F=0$
(solid line), 4.0 (dash line), 8.0 (dotted line), and 10.0 kV/cm (dash-dotted
line).} \label{FIG:ps cqds}
\end{center}
\end{figure}

Including the exchange interaction $V_{\mathrm{ex}}$, the exciton ground states
are split into bright and dark doublets. The polarization splittings of bright
doublets in single InGaAs QDs are mainly determined by the anisotropic shape
and the exchange interaction.~\cite{Bayer,Takagahara} However, the behavior of
the polarization splitting in CQDs is greatly different from that of single
QDs. Fig.~\ref{FIG:ps cqds}(a) shows the numerically-calculated polarization
splitting of the exciton ground state in CQDs as a function of the in-plane
electric field alone the $x$-axis for $d=3.5, 4.0, 4.5$, and $5.0$ nm. The
behavior of the polarization splitting is asymmetric since two QDs are
nonidentical. At $F=0$, the polarization splitting increases as $d$ decreases,
from $-30$ $\mu$eV for $d=5.0$ nm to $+8$ $\mu$eV for $d=3.5$ nm. For a given
$d$, the polarization splitting increases initially as $F$ is applied, but
decreases when $F$ becomes larger. Interestingly, for $d$ larger than about 3.7
nm, the polarization splitting could be tuned to zero by the electric field at
$F_0$. For instance, $F_0=3.2$ or $-6.6$ kV/cm for $d=4.0$ nm, $F_0=6.1$ or
$-8.7$ kV/cm for $d=4.5$ nm, and $F_0=7.3$ or $-9.5$ kV/cm for $d=5.0$ nm. At
the same time, the exciton ground states at $F_0$ remain optically active. For
examples, the ratio of the oscillator strength of the exciton ground state at
positive $F_0$ to that at zero electric field is 1.02, 0.988, and 0.863 for
$d=4.0$, 4.5, and 5.0 nm, respectively. However, for $d>5.0$ nm the ratio would
be largely reduced, which means that the exciton ground states might be less
efficient in the spontaneous emission of photons. It can be concluded that the
best range of $d$ for the effective tuning of the polarization splitting is
$3.7\sim5.0$ nm.

In order to understand the underlying mechanisms, we introduce a model
Hamiltonian on the basis $|00\rangle$, $|01\rangle$, $|10\rangle$, and
$|11\rangle$
\begin{eqnarray}\label{44matrix}
\left( \begin{array}{cccc}
|00\rangle&|01\rangle&|10\rangle&|11\rangle\\
E_{00}&t_h&t_e&0\\
t_h&E_{01}-eF(d+a_0)&0&t_e\\
t_e&0 &E_{10}+eF(d+a_0)&t_h\\
0&t_e&t_h&E_{11}
\end{array}\right)
\end{eqnarray}
where $|ij\rangle$ indicates that the electron and hole localize in the $i$th
and $j$th dot, respectively, $t_e$ ($t_h$) is the tunneling energy of the
electron (hole), $E_{ij}$ is the energy of the exciton with the electron and
hole localized in the $i$th and $j$th dot, respectively, and $F$ is the
electric field. Inter-dot Stark effect is approximated by $eF(d+a_0)$ while
intra-dot Stark effect is neglected. Due to attractive Coulomb interaction
between the electron and hole, $E_{01}$ and $E_{10}$ are larger than $E_{00}$
and $E_{11}$. The ground state has the following form as
$\psi_1=c_0|00\rangle+c_1|11\rangle+c_2|01\rangle+c_3|10\rangle$. At zero
electric field, main components of the ground state $\psi_1$ and the second
state $\psi_2$ are $|00\rangle$ and $|11\rangle$ while those of the third state
$\psi_3$ and fourth state $\psi_4$ are $|01\rangle$ and $|10\rangle$. As
electric field increases, diagonal element $[E_{01}-eF(d+a_0)]$ of $|01\rangle$
decreases and the couplings between state $|01\rangle$ and intra-dot states
($|00\rangle$ and $|11\rangle$) are enhanced. Therefore, significant Stark
effect and anticrossing can been seen in the energy spectrum as shown in
Fig.~\ref{FIG:fs cqds}(a).

In the first order approximation, formula of the polarization splitting of
exciton ground state could be obtained using the basis introduced above. It is
as follows
\begin{eqnarray}\label{ps formula}
\delta=2\langle\psi_1,+1|V_{\mathrm{ex}}|\psi_1,-1\rangle=2(\lambda_{\mathrm{intra}}+\lambda_{\mathrm{inter}})
\end{eqnarray}
where $\lambda_{\mathrm{intra}}=\sum_{i=0,1}c^2_i\langle
ii,+1|V_{\mathrm{ex}}|ii,-1\rangle$ is the electron-hole exchange interaction
within individual QDs, $\lambda_{\mathrm{inter}}=\sum_{i\neq
j(=0,1)}c_ic_j\langle ii,+1|V_{\mathrm{ex}}|jj,-1\rangle$ is the exchange
interaction between two QDs, and $|ii,s\rangle$ is the wave function of exciton
state $|ii\rangle$ with exciton spin $z$-component $s=+1$ or $-1$. The
polarization splitting of exciton ground state in CQDs at zero electric field
is significantly affected by the inter-dot coupling as shown in
Fig.~\ref{FIG:ps cqds}(a). Due to anisotropy of $V_{\mathrm{ex}}$,
$\lambda_{\mathrm{intra}}<0$ $(>0)$ for $a_i<b_i$ $(a_i>b_i)$ as demonstrated
in Ref.~\onlinecite{Ivchenko} while $\lambda_{\mathrm{inter}}>0$ $(<0)$ for the
two dots aligned alone the $x$-axis ($y$-axis). In order to effectively tune
the exciton polarization splitting, $a_i$ and $b_i$ should be chosen to make
$\lambda_{\mathrm{intra}}$ and $\lambda_{\mathrm{inter}}$ have opposite sign.
Here, we have two nonidentical dots with $a_i<b_i$ aligned alone the $x$-axis.
As the inter-dot separation $d$ is reduced, both of the inter- and intra-dot
part of the exchange-interaction are greatly changed by the tunnel-coupling,
and thus the intra-dot part is largely compensated or even exceeded by the
inter-dot part. This is the reason why we use CQDs instead of single QD to
effectively tune the polarization splitting.


It is necessary to examine the variation of the components in the exciton
ground state under the electric field. Taking $F$ at positive direction as
example, Fig.~\ref{FIG:ps cqds}(b) shows numerically-calculated
$\psi(r_e,r_h=r_e)$ of the exciton ground state alone $y=0$ axis in CQDs with
$d=4.0$ nm for several values of electric field. It can be found that as $F$
increases the component of $\psi(r_e,r_h=r_e)$ in $1$th dot (proportional to
component $|11\rangle$) is greatly reduced while that in $0$th dot
(proportional to component $|00\rangle$) is not much changed until larger
electric field. Equivalently speaking, $c_1$ of the ground state $\psi_1$ is
largely reduced due to the coupling between states $|01\rangle$ and
$|11\rangle$ while $c_0$ remains nearly unchanged before the anticrossing of
the energy levels. $|\lambda_{\mathrm{intra}}|$ which is proportional to
$c^2_1$ decreases faster than $\lambda_{\mathrm{inter}}$ which is proportional
to $c_1$. Therefore, for proper $d$ with $\delta<0$ at $F=0$, $|\delta|$ is
initially decreased by $F$ and could be tuned to zero at $F_0$ where
$\lambda_{\mathrm{inter}}$ is completely compensated by
$\lambda_{\mathrm{intra}}$. After the anticrossing as shown in Fig.~\ref{FIG:fs
cqds}(a), $c_0$ and $c_1$ both are greatly reduced and the exciton ground state
would become optically inactive inter-dot state $|01\rangle$. Therefore the
polarization splitting finally approaches zero at larger electric field as
shown in Fig.~\ref{FIG:ps cqds}(a).

In order to better understand the effective tuning in CQDs, we also calculate
the polarization splitting of the exciton ground state in a single QD with
$a=16$ nm and $b=20$ nm, under an in-plane electric field alone $x$-axis. The
polarization splitting under zero electric field is about $-66 \mu$eV, which is
in good agreement with the experimental results of QDs with similar
sizes.~\cite{Bayer} As shown in Fig.~\ref{FIG:ps cqds}(a), the absolute value
of the polarization splitting in a single QD decreases as the electric field
increase, which is consistent with the experimental result.~\cite{Kowalik} The
electric field reduces the overlapping between the electron and hole, and thus
the exchange interaction as well as the polarization splitting might be
weakened. However, the reduction of the polarization splitting is small even
for $F>16$ kV/cm. Thus it is difficult to effectively tune the splitting in a
single QD by the electric field.

Compared with single QDs, energy levels of exciton in CQDs under the electric
field would exhibit stronger Stark effect and there are significant
anticrossing between the inter- and intra-dot exciton states. It would greatly
modify the exciton ground states and therefore the electron-hole exchange
interaction. Moreover, the intra-dot exchange interaction in CQDs at zero
electric field is already largely compensated by the inter-dot exchange
interaction at proper inter-dot separations. In that case, polarization
splitting in CQDs could be effectively tuned to zero by the in-plane electric
field and at the same time the exciton states remain optically active.
Interestingly, as $d$ approaches zero or infinity, CQDs will become a single
larger QD or two isolated QDs, respectively. It will be again difficult for the
effective tuning of the exciton polarization splitting by the electric field.

In summary, we study the polarization splitting of exciton ground state in two
laterally CQDs, and find that it could be effectively tuned to zero by the
in-plane electric field as well as the structural design, which overcomes the
deficiency in a single QD. Our study will be helpful and interesting for the
research on the QDs-based entangled-photon source.

Financial supports from NSF-China (Grant No. 10574077), the ``863" Programme
(No. 2006AA03Z0404) and MOST Programme (No. 2006CB0L0601) of China are
gratefully acknowledged.


\end{document}